\documentclass[prl,twocolumn,showpacs]{revtex4}
\usepackage{amsmath}
\usepackage{bm}
\usepackage{latexsym}
\usepackage{epsfig}

\newcommand{\x}{{\bf x}}
\newcommand{\p}{{\bf p}}

\begin{document}

\title{Perturbation spectrum in inflation with cutoff}

\author{Achim Kempf}
\email[]{akempf@math.uwaterloo.ca}
\altaffiliation{New affiliation from Aug.1 2001: Dept. of
Applied Mathematics, University of Waterloo, Waterloo, Ontario, N2L
3G1, Canada}
\affiliation{Institute for Fundamental Theory, Departments
 of Physics and Mathematics,
University of Florida, Gainesville, FL 32611, USA}
\author{Jens C. Niemeyer}
\email[]{jcn@mpa-garching.mpg.de}
\affiliation{Max-Planck-Institut f\"ur Astrophysik,
Karl-Schwarzschild-Str. 1, D-85748 Garching, Germany}

\begin{abstract}
It has 
been pointed out that the perturbation spectrum predicted by inflation
may be sensitive to a natural ultraviolet cutoff, thus potentially
providing an experimentally accessible window to aspects of Planck
scale physics.   
A priori, a natural ultraviolet cutoff could take any form, but a
fairly general classification of possible Planck scale cutoffs
has been given. One of those categorized cutoffs, also appearing in
various studies of quantum gravity and string theory, has recently
been implemented into the standard inflationary scenario.
Here, we continue this approach by investigating its effects on the
predicted perturbation spectrum. We find that 
the size of the effect depends sensitively on the scale separation
between cutoff and horizon during inflation.
\end{abstract}


\maketitle

\section{Introduction}

During the inflationary phase of the very early universe (see
e.g. \cite{LL00} for an overview and references) space-time is assumed to 
expand in a quasi-exponential fashion.
Quantum fluctuations of the inflaton field are continuously redshifted
until their wavelength equals the physical horizon distance, whereupon
they become ``frozen'' until they re-enter the Hubble volume during
the ensuing radiation or matter dominated epochs. These
fluctuations are thought to be
responsible for seeding the temperature fluctuations of the cosmic
microwave background radiation (CMBR) and the gravitational clustering
of matter, whose statistical properties may therefore provide a window into the
realm of high-energy physics. 

Crucially, in the case of a sufficiently long period of inflation,
all of the scales of cosmological interest today correspond to
wavelengths below the Planck length early on when the initial
conditions are prescribed. Therefore, inspired by similar studies in the context
of Hawking radiation \cite{U81,J91,U95,BMPe95,CJ96,BGLe99,SS00}, a series of papers
\cite{MB00,N00,KG00,NP01} has investigated the sensitivity of the
predictions of inflationary scenarios with respect to changes of
trans-Planckian physics. Those studies encoded transplanckian physics
in a simple way as nonlinearities of
the dispersion relation of the Fourier mode functions (see also
\cite{MBP01} for a different application of this ansatz). 

Since linearity
of the field and hence Gaussianity of the fluctuations remains
unchanged, the potential consequences of such modifications are limited
to a possible scale dependence of the power spectrum and 
a possible change in its overall amplitude.
It was shown \cite{NP01} that under rather general
conditions on the dispersion relation no observable effects can be
expected, although Ref.~\cite{MB00} reaches a 
somewhat different conclusion.
However, those studies suffer from fundamental limitations. First of
all, with the exception of Ref.~\cite{KG00} all of the employed
dispersion relations were chosen {\it ad hoc} so as to provide bounds
on the frequency, wavelength or both without reference to an underlying
theory. More importantly, the question of mode
generation, i.e.~how each semiclassical quantum field degree of
freedom emerges out of the Planck regime, has not been addressed. 

In contrast, Ref.~\cite{K00} proposes a scenario where the UV cutoff is
provided by a modified quantum mechanical commutation relation that 
limits the experimentally attainable resolution of small spatial
distances \footnote{The same approach was taken in the black hole
context in Ref.~\cite{BGLe99}.}. This UV cutoff is one of very few
types of short-distance structures that appear in the classification presented in
\cite{K98}, which applies to all quantum gravity theories that effectively
represent each dimension by a linear operator. Indeed, corresponding 
short-distance uncertainty relations of this kind have appeared
in various studies of quantum gravity and string theory, see e.g.\cite{W96}. 
In Ref.~\cite{K00} this short distance cutoff has been implemented into the
theory of a minimally 
coupled scalar field living in an expanding Friedmann-Robertson-Walker
(FRW) background and it has been shown how the 
decoupling degrees of freedom are continuously generated
dynamically at the time of their ``Planck scale crossing''. Here, we
aim to extend the analysis of \cite{K00} by estimating the magnitude of
any corrections to the standard predictions for the
statistical distribution of inflationary perturbations arising from
the modified short-distance behavior.

The approach in Ref.~\cite{K00} which we will follow here utilizes that,
as has been shown in \cite{K94}, the quantum gravity and stringy uncertainty relation
cutoff (see e.g. \cite{W96}) can be modeled by corrections to the commutation relations
\begin{equation}
\label{nccr}
[\x,\p]=i (1+\beta \p^2)
\end{equation}
and its higher dimensional generalizations. 
It is easy to check that such correction terms give rise to a lower bound $\Delta x_{min} =
\sqrt{\beta}$ for distance measurements. 
The form of these correction terms
is unique to first order in $\beta$. Correspondingly, the signature of 
the first order effects of this type of natural cutoff should be unique when moving 
up from low energies, see \cite{K97}.

Hilbert space
representations of relations of the type of Eq.~(\ref{nccr}) are given by introducing an
auxiliary variable $\rho$ which is essentially the momentum variable $p$ but differs
from it at small distances, i.e. at distances close to $\Delta x_{min}$. While this is
initially a quantum mechanical structure, it can be implemented into
quantum field theory, see Ref.~\cite{K00}. Within the scalar quantum
field theory on an inflationary background
as defined in Ref.~\cite{K00} one finds that, interestingly, those variables, $\tilde{k}$, in which
the mode equations decouple, no longer strictly coincide with the comoving momentum 
variables, $k$, although they do of course approximately coincide for small $k$, i.e. for
large distances. Conversely, this means that the comoving momentum modes 
now decouple only 
when they have grown to large proper distances and that the comoving momentum modes
do couple initially when they emerge from the cutoff scale. For the quantum theory of the
actual mode creation  mechanism, see Ref.~\cite{K00}.
Explicitly, one obtains within this framework the following mode equation for
the decoupling $\tilde k$-modes:
\begin{equation}
\label{eom1}
\phi_{\tilde{k}}^{\prime\prime} + \frac{\nu^\prime}{\nu} 
\phi_{\tilde{k}}^\prime +\left(
\mu
 - 3 \left(\frac{a^\prime}{a}\right)^\prime 
-9\left(\frac{a^\prime}{a}\right)^2
-\frac{3 a^\prime \nu^\prime}{a \nu}
\right)\phi_{\tilde{k}} = 0\,\,.
\end{equation}
Here, $a$ is the scale factor of the FRW line element and we defined the functions
\begin{eqnarray}
\label{munu}
\mu(\eta,\tilde{k}) & := & 
-\frac{a^2 \mbox{plog}(-\beta \tilde{k}^2/a^2)}{
\beta\left(1+\mbox{plog}\left(-\beta \tilde{k}^2/a^2\right)\right)^2}
\\
\nu(\eta,\tilde{k}) & := & 
\frac{e^{-\frac{3}{2}\mbox{plog}\left(-\beta \tilde{k}^2/a^2\right)}}{
a^4 \left(1+\mbox{plog}\left(-\beta \tilde{k}^2/a^2\right)\right)}
\end{eqnarray}
that utilize the ``product log'' plog, which is the inverse of the
function $x \rightarrow x e^x$.  
The solutions are automatically defined only from a finite value of
$\eta$, i.e. every mode possesses its ``creation time". It is the time when,
in terms of proper distances, 
the mode equals the size of the cutoff scale, i.e.
it is the conformal time $\eta_c$ defined implicitly by
\begin{equation}
\label{etac}
a(\eta_c) = \tilde k \sqrt{e \beta} \sim \tilde k \, \Delta x_{min}\,\,.
\end{equation}
At the creation time, the differential equation possesses 
what is called an irregular singular point. To see this, note that
the function plog which enters the differential equation
through the functions $\mu$ and $\nu$ is not analytic at the creation time.
Below, we will further discuss possible implications for the choice of initial conditions
and therefore for the choice -- or possible uniqueness -- of the initial vacuum.

All physical observables in our model universe can be expressed 
in terms of $\tilde k$ instead of the usual Fourier variable $k$. This
argument applies also to the transfer function $T(t, \tilde k)$ which
relates today's observable perturbations to the horizon crossing
amplitude of $\phi_{\tilde k}$, provided that the perturbation
amplitudes are measured as a function of $\tilde k$. In practice,
these measurements are carried out on cosmological scales where
$\tilde k = k$ to extremely good accuracy, so we do not expect any
consequences from the re-labelling of physical observables such as
the angular size distribution of CMBR fluctuations. In other words, any
statement about the scale dependence and Gaussianity of the horizon crossing
amplitudes of $\phi_{\tilde k}$ translate into corresponding
statements about CMBR fluctuations, at least to the same extent as in the standard
theory. Let us note, however, that this would not be true if the cutoff had
different properties for different fields, e.g. if linear metric fluctuations
behave differently on small scales than the inflaton field. The
following analysis assumes that this is not the case \footnote{At
horizon crossing of the mode $\tilde k$, i.e. when $\tilde k \approx a H$, we also note
that $\tilde k^2$ and $k^2$ differ only by the constant factor $-
\sigma^{-2} \mbox{plog}(-\sigma^2)$, independent of $k$, where
$\sigma$ is defined in Eq.~(\ref{sigdef}).}.

Eq.~(\ref{eom1}) is linear in $\phi_{\tilde k}$ so that Gaussianity of
the distribution of fluctuations in $\tilde{k}$-space  is protected.
Consequently, we expect no deviations from Gaussianity owing to the proposed modifications of the
short-distance behavior. We can therefore restrict attention to examining possible new effects
on the scale dependence and overall amplitude of the power spectrum.

\section{Analysis in oscillator variables}
\label{oscillator}

It turns out to be very convenient to change from the field variables used in  
Ref.~\cite{K00} to slightly new variables defined by:
\begin{equation}
\label{phidef}
\varphi_{\tilde k} \equiv \nu^{1/2} \phi_{\tilde k}\,\,.
\end{equation}
Indeed, while the mode equation  Eq.~(\ref{eom1})
in terms of the original field $\phi$ is of the type of 
a harmonic oscillator with friction, there is no friction term in the
mode equation when written in terms of the new variable $\varphi$:
\begin{equation}
\label{eom2}
\varphi_{\tilde k}'' + \omega^2(\eta) \varphi_{\tilde k} = 0
\end{equation}
where $\omega(\eta)$ obeys  the time dependent, nonlinear dispersion relation
\begin{equation}
\label{omega}
\omega^2(\eta) = \mu - 6 \left(\frac{a'}{a}\right)^2 + \left(\frac{\nu'}{2 \nu}\right)^2 -
\frac{3 (a' \nu' + a'' \nu)}{a \nu} - \frac{\nu''}{2 \nu}\,\,.
\end{equation}
The Wronskian condition from Ref.~\cite{K00} now also simplifies to
\begin{equation}
\label{wronski}
\varphi_{\tilde k} {\varphi_{\tilde k}^\ast}' - \varphi_{\tilde k}' \varphi_{\tilde k}^\ast = i
\end{equation}
as usual. Note also that if we denote the standard field mode with a
vanishing minimum position uncertainty as $\chi_{\tilde k} = \varphi_{\tilde k}(\beta \to
0)$,  
we obtain the usual equation of motion for the $\tilde k$ mode of a free, minimally 
coupled scalar field in an expanding FRW space-time, where $\chi$ is in the conventions of e.g.
\cite{BD84}
 \footnote{Of course, also
$\tilde k = k$ for $\beta \to 0$. However, for the reasons explained above we
prefer to label everything in terms of $\tilde k$ in order to avoid
discussing the $\tilde k \to k$-map.}:
\begin{equation}
\label{standardmode}
\chi_{\tilde k}'' + \omega_0^2 \chi_{\tilde k} = 0
\end{equation}
with
\begin{equation}
\label{omnull}
\omega_0^2 = {\tilde k}^2 - \frac{a''(\eta)}{a(\eta)}\,\,.
\end{equation}

Again, there is the question of initial conditions for
$\varphi_{\tilde k}$ that determine the vacuum for $\hat
\phi$. Ideally, regularity arguments at the irregular singular point
of the mode equation, encountered at the creation time $\eta_c$ for
each mode, should fix the choice.  We do not have a definite answer at
this point, but asymptotic methods will shed some light on the
situation. Some indications of vacuum fixation by regularity arguments
are sketched in the Conclusions. Indeed, a solution of the singularity
problem is not strictly necessary for the present analysis. It will be
shown below that the evolution of $\varphi_{\tilde k}$ is essentially
adiabatic from a certain time $\eta_i$ onwards.
The state of $\hat \phi$ at $\eta \ge \eta_i$ can be determined by
consistency arguments to be the adiabatic vacuum (e.g.,\cite{BD84})
\begin{equation}
\label{vac}
\varphi_{\tilde k}(\eta) = \frac{1}{\sqrt{2 \omega(\eta)}}
 \exp\left(-i  \int^\eta_{\eta_i} \omega(\tilde \eta) d\tilde
 \eta\right)\,\,,
\end{equation}
where the normalization follows from Eq.~(\ref{wronski}). This is
because, as argued in Refs.~\cite{NP01,T00}, any small deviation from the 
adiabatic vacuum close to the Planck scale would likely suppress
inflation altogether due to back-reaction of the energy density
contained in $\varphi_{\tilde k}$ on the cosmic expansion. In order 
to be consistent with the assumptions of Ref.~\cite{K00} (i.e.,
negligible back-reaction), any admissible initial condition needs to
converge to the adiabatic vacuum as soon as the latter is well
defined.

\section{Adiabatic analysis}
\label{adiabatic}

Eq.~(\ref{eom2}) belongs to the class of harmonic oscillator equations
featuring a dispersion relation that is asymptotically linear for
small physical wavenumbers but becomes nonlinear at high wavenumbers
(small wavelengths). In the context of cosmology, such systems were
investigated in Refs.~\cite{MB00,N00,NP01}, and in
the framework of Hawking radiation many times before (see \cite{J00}
for references). Unlike in the above references, where the dispersion
relation was typically tailored {\it ad hoc} to fit some desired
shape, Eq.~(\ref{omega}) followed directly from a general study of 
realistic short-distance structures of space-time and may therefore perhaps 
be considered more fundamental (see also Ref.~\cite{KG00} for a similar
approach). 

It is useful to express the separation between the cutoff scale
(here parameterized by $\beta^{1/2}$) and the inflationary horizon
scale in terms of the dimensionless parameter $\sigma$ defined as
\begin{equation}
\label{sigdef}
\sigma \equiv \beta^{1/2} H\,\,.
\end{equation}
If $\sqrt{\beta} \sim \Delta x_{min}$ is identified with the Planck
length, the amplitude of temperature fluctuations of the cosmic
microwave background indicates that $\sigma \sim 10^{-5}$ at the time
when the presently observable scales left the horizon during
inflation. 

In order to generalize the notion of ``horizon crossing'' to our
non-standard equation of motion, we Taylor-expand
Eq.~(\ref{omega}) around $\sigma =0$ and find that $\omega(\eta)^2 =
\omega_0^2 + O(\sigma^2)$. Correspondingly, the usual definition of
horizon crossing in terms of $\tilde k = a H$ is valid to within the
same accuracy.

We are interested in sources of deviation from the
standard (i.e., $\beta \to 0$) result for the scale dependence and
overall normalization of the horizon crossing amplitude of
$\phi_{\tilde k}$. Following Sec.~(\ref{oscillator}) and recognizing
that $\varphi_{\tilde k} = \phi_{\tilde k} +  O(\sigma^2)$ at horizon
crossing, we need to compare the amplitudes of 
$\varphi_{\tilde k}$ and $\chi_{\tilde k}$ at the horizon crossing
time $\eta_h$, which is when $\tilde k \approx a(\eta_h)
H$
 \footnote{Equivalently, we could compare $\phi_{\tilde k}$ and $a^2
\chi_{\tilde k}$, as $\nu \to a^{-4}$ for $\beta \to 0$.}.

%\begin{figure}[t]
%\epsfxsize=0.45\textwidth
%\epsfbox{fuzzy_C.eps}
%\caption{\label{f1} Logarithm of the adiabaticity parameter ${\cal C}$
%for $\sigma = 10^{-5}$ and $\tilde k = 1$, beginning at $\eta_i =
%\eta_c(1 +\epsilon)$, for $\epsilon=10^{-5}$.
%The sudden rise of ${\cal C}$  for $\eta \to 0$
%signals the usual onset of non-adiabaticity due to cosmic expansion.}
%\end{figure}

One possible signature of the cutoff in the spectrum is due to
non-adiabatic particle production during the evolution from $\eta_i$
to $\eta_h$, which may give rise to a modulation of $\varphi_{\tilde
k}(\eta_h)$ around the amplitude predicted for $\beta \to 0$
\cite{NP01}. This may, in turn, be reflected by a breaking of scale invariance
of the perturbation power spectrum. The {\it relative} magnitude of
this effect, denoted in 
Ref.~\cite{NP01} as $\beta_k$, can be shown to be bounded by the
maximum of the adiabaticity parameter
\begin{equation}
\label{cdef}
{\cal C}(\eta) = \left|\frac{\omega'(\eta)}{\omega^2(\eta)}\right|\,\,.
\end{equation}
If ${\cal C} \lesssim 1$, the usual notions of semiclassical quantum field
theory in nonstationary space-times apply, with the adiabatic vacuum,
Eq.(\ref{vac}), serving as a natural ground state. One finds numerically
that ${\cal C} \approx 1$ for $\eta= \eta_i$ and drops to
neglibigle levels afterwards, where
\begin{equation}
\eta_i \approx \eta_c (1 + \sigma^2)
\end{equation}
for $\sigma$ ranging from $10^{-7}$ to $0.1$. Perhaps not surprisingly,
the beginning of adiabaticity approaches the initial singularity
arbitrarily closely if cosmic expansion becomes sufficiently
slow. Conversely, any bound on non-adiabatic particle production
due to the cutoff derived from inflation is stronger than the
equivalent bound from cosmic expansion today.
Furthermore, if one defines a natural time scale for the evolution of
$\varphi_{\tilde k}$ at the time $\eta_i$ as $\tau(\eta_i) \sim
1/\omega(\eta_i)$, one can check numerically that $\eta_i \approx 0.75
\tau$. In other words, the ``non-adiabatic epoch'' following the
Planck scale crossing of each mode lasts about as long as the typical
evolution time scale of the mode itself. Whatever physics determines
this phase remains, for the time being, unknown. However, as
argued in Sec.~(\ref{oscillator}), self-consistency demands the
solution to converge onto the adiabatic vacuum as soon as it is well
defined (i.e., as soon as ${\cal C} \ll 1$) \cite{NP01,T00} and this
is the case for all $\eta \gtrsim \eta_i$.

Having shown that scale invariance is preserved if $\sigma$ is small,
we need to consider the overall amplitude of the power spectrum.
Taking the adiabatic solution Eq.~(\ref{vac}) as a reasonable
approximation to the exact functions $\varphi_{\tilde k}(\eta)$ and
$\chi_{\tilde k}(\eta)$ on length scales larger than the cutoff but
smaller than the horizon 
scale (where expansion violates adiabaticity), i.e. for times $\eta_i
\ll \eta \ll \eta_h$, one finds that 
\begin{equation}
D(\eta) \equiv \frac{\varphi_{\tilde k}(\eta)}{\chi_{\tilde k}(\eta)} =
\left(\frac{\tilde k}{\omega(\eta)}\right)^{1/2}\,\,.
\end{equation}
A good estimate for the impact of the nonlinear dispersion relation
on the amplitude of the power spectrum is obtained by noting that
this expression for $D(\eta)$ remains approximately valid until
$\eta_h$, since 
cosmic expansion affects both solutions in roughly the same
way. It is readily verified in this case that $D(\eta_h) = 1 +
O(\sigma^2)$. Hence, the impact of the cutoff on the perturbation 
spectrum depends crucially on the separation between the cutoff and the Hubble
scale, being negligible if $\sigma \ll 1$.

\section{Scaling analysis}
\label{scaling}

The scaling behavior of the perturbation  spectrum can also be
investigated by studying the scaling behavior of the wave equation
Eq.~(\ref{eom1}). 
Let us begin by considering the case of an exactly de Sitter type
expansion.  In this case, we expect that time translation invariance
is broken neither  by our introduction of a cutoff nor by the
background  expansion. We therefore expect a scale invariant
perturbation spectrum.

Indeed, we first observe that if  $\phi_{\tilde{k}}(\eta)$ is a
solution to the $\tilde{k}$ mode equation and $r$ is any arbitrary
positive number then  $\phi_{\tilde{k}}(r \eta)$ is a solution of the
mode equation for the mode $r \tilde{k}$.  This is straightforward to
verify and it is of course also true for the usual inflationary
scenario without a cutoff.

The solutions $\phi_{r \tilde{k}}(\eta)$ that are obtained in this way
by scaling the solution $\phi_{\tilde{k}}(\eta)$ all obey of course the
same initial conditions. We can also conclude that if $\eta$ is a
special time for the solution $\phi_{\tilde{k}}$, then,
correspondingly,  $\eta/r$ is a special time for  the solution
$\phi_{r \tilde{k}}$. For example, if we denote the  creation and the
horizon  crossing times of the mode $\phi_{\tilde{k}}$ by  $\eta_c$
and $\eta_h$, then  the  mode $\phi_{r \tilde{k}}(\eta)$ possesses the
creation and the horizon crossing times $\eta_c(r \tilde{k})=
\eta_c/r$ and $\eta_h(r\tilde{k})= \eta_h/r$.

Let us further assume that the solution $\phi_{\tilde{k}}(\eta)$
is normalized with respect to the Wronskian condition. We also need
that all the solutions  $\phi_{r \tilde{k}}(\eta)$  are normalized
with respect to the Wronskian condition for the respective  $r
\tilde{k}$ modes. As is straightforward to verify, the ansatz
\begin{equation}
\phi_{r \tilde{k}}(\eta) = N(r) \phi_{\tilde{k}}(r \eta)
\end{equation}
yields
\begin{equation}
N(r) = r^{3/2}
\end{equation}
so that $\phi_{r \tilde{k}}(\eta) = r^{3/2} \phi_{\tilde{k}}(r \eta)$,
and therefore:
\begin{equation}
\phi_{r \tilde{k}}(\eta/r) = r^{3/2} \phi_{\tilde{k}}(\eta)
\end{equation}
Choosing for $\eta$ the horizon crossing time of the $\tilde{k}$ mode
we now obtain how the horizon crossing amplitude scales when scaling
the decoupling momentum
\begin{equation}
\phi_{r \tilde{k}}(\eta_h(r\tilde{k})) = r^{3/2}
\phi_{\tilde{k}}(\eta_h)
\label{e1}
\end{equation}
which means:
\begin{equation}
\phi_{r \tilde{k}}(\mbox{horizon crossing}) \sim r^{3/2}
\end{equation}
In order to make contact with the conventions in the literature, let
us now  recall that, usually, field variables $\psi(\eta,k)$ in
comoving momenta $k$ are obtained by first scaling from proper
position coordinates to comoving position coordinates and then,
second, by Fourier transforming to the comoving momentum.  In
\cite{K00}, however,  we obtained fields $\phi(\eta,k)$ over comoving
momenta $k$  by first Fourier transforming from proper positions to
proper momenta and then, second, by scaling to comoving
momenta.
However, scaling and Fourier transforming do not commute.  As
a consequence, as is readily verified: 
\begin{equation}
\phi(\eta,k) = a^{3} \psi(\eta,k)
\end{equation}
and in the de Sitter case:
\begin{equation}
\phi(\eta,k) = -\frac{H^3}{\eta^3} \psi(\eta,k)
\end{equation}
As far as present day observations of cosmological scales are concerned,
the distinction between comoving and decoupling momenta does not matter and we
therefore obtain from Eq.\ref{e1}:
\begin{equation}
\psi(\eta_h/r,r k) = r^{-3/2} \psi(\eta_h,k)
\end{equation}
We therefore finally obtain for the fields over comoving momenta as
conventionally defined the scaling behavior of the horizon crossing
amplitude
\begin{equation}
\psi(\mbox{horizon crossing},r k) \sim r^{-3/2}
\end{equation}
which yields indeed the usual scale invariant spectrum: 
\begin{equation}
\langle 0\vert \psi^\dagger(\mbox{horizon crossing},  r
k)\psi(\mbox{horizon crossing},r k)\vert 0\rangle \sim r^{-3}
\end{equation}
Indeed, this was to be expected because neither the background de Sitter space,
nor our introduction of a cutoff, nor the choices of initial conditions
(all solutions being obtained from another by mere scaling) 
 broke time translation invariance.

On the other hand, in the case of a non-de Sitter background, the
spectrum is of course not scale invariant. In the presence of our
cutoff we  will then obtain additional scale invariance breaking
effects on the spectrum, because of the new cutoff dependent terms in
the wave equation.

\section{Conclusions}
We investigated the signature of the cutoff in the perturbation
spectrum from two perspectives, and in both cases we did 
not need to solve the
mode equation explicitly. The adiabatic treatment in
Sec.~(\ref{adiabatic}) is based on the fact that in order to be
consistent with inflation, each mode needs to be
in the adiabatic vacuum shortly after the mode is created, whereas the
scaling analysis of Sec.~(\ref{scaling}) utilizes that  the wave
equation scales trivially and that there is also no reason for  the
(still unknown) initial conditions to break the (almost)
time-translation invariance of the background space-time. Both
approaches show that the resulting fluctuation power spectrum is
indeed scale invariant if the background space-time is de
Sitter. The adiabatic analysis,
in addition, shows that any corrections of the overall amplitude are
at most of order $\sigma^2$, where $\sigma$ is the ratio of the
horizon scale and the minimum spatial resolution $\Delta x_{min}$
admitted by the commutation relation Eq.~(\ref{nccr}). 

While a detailed analysis in the framework of slow-roll
inflation would be desirable, we expect the following first order
effect in a more general inflationary space-time. For small $\sigma$, the
dispersion relation, Eq.~(\ref{omega}), can be approximated by
\begin{equation}
\omega^2 = \omega_0^2 + B\,\sigma^2 + O(\sigma^4)\,\,,
\end{equation}
with
\begin{equation}
B = \frac{3 k^4}{a^2 H^2} + \frac{5 k^2}{2 H^2}\left(\frac{a''}{a^3} -
H^2\right)\,\,.
\end{equation}
During inflation, the pressure is roughly equal to the negative energy
density, so that the Friedmann equations yield $a''/a^3 \approx 2 H^2$.
Therefore, $B > 0$ which implies that the dispersion relation is
superluminal in the region of interest. Owing to the normalization in
Eq.~\ref{vac} (cf.~Refs.\cite{N00,NP01}), the 
perturbation amplitude drops more quickly than usual as $H$ declines 
with time, giving rise to additional reddening of the
spectrum. Evidently, this effect vanishes as either the slow-roll
parameter or $\sigma$ go to zero.

The scale $\Delta x_{min}$ at which a natural
ultraviolet cutoff sets in could be as small as the $3+1$ dimensional
Planck scale of $10^{-35}$ meters, but the natural short distance
cutoff scale may well be larger, as could be the case e.g. in string
theory and theories of large extra dimensions.
Evidently, the  signature of the cutoff in the CMBR would increase if
the cutoff scale were larger than the Planck length during
inflation. On the other hand, both the scale dependence and the 
amplitude of the power spectrum are very sensitively dependent on the
details of the inflaton potential. Only after a concrete model for
inflation has been specified one can derive an upper bound on
$\Delta x_{min}$ from observations.

In particular, if we assume conventional slow roll inflation
with the inflaton
coupling fine-tuned such as to obtain the observed amplitude of the 
CMBR perturbations, then a cutoff $\Delta
x_{min}$ at the Planck length suggests
$\sigma \sim 10^{-5}$. The cutoff induced corrections to the perturbation 
spectrum would then be neglible, i.e. the conventional scenario with its
parameters fine-tuned as usual is observationally consistent with
a cutoff $\Delta x_{min}$ at the Planck scale. 

On the other hand, we may view $\Delta x_{min}$ simply as a new 
free parameter in model building. For example, it might be possible
to gain some more freedom in choosing the potential --
for the prize of having to fine-tune $\Delta x_{min}$. 

An interesting technical question remains:  We have not shown
how or even if the decoupling modes evolve into the adiabatic vacuum
from some natural initial conditions at the singularity. Two
possibilities can be imagined: either there exists a symmetry or
regularity condition that uniquely specifies initial conditions that
later evolve into the adiabatic solution. In this case the discussion
in Sec.~(\ref{adiabatic}) applies.

Or, alternatively, the modes are generally created in a highly excited
state as seen from the point of view of a comoving particle
detector. This case would be inconsistent \cite{NP01,T00}  with the
assumption of slow-roll inflation made at the onset of
Ref.~\cite{K00}, indicating that the combination of
short-distance uncertainty of the kind described by Eq.~(\ref{nccr}) 
and inflation is not, in general, self-consistent.

We will conclude with some speculative ideas about the first
possibility for the initial conditions at the singularity. Starting
with the original equation of motion, Eq.~(\ref{eom1}), expanding the
coefficients around $\eta = \eta_c$, and shifting the origin of the
time coordinate to $\eta_c$, one obtains a differential equation of the
form
\begin{equation}
\phi_{\tilde k}'' - \frac{1}{2 \eta} \phi_{\tilde k}' + \frac{A}{\eta}
\phi_{\tilde k} = 0
\end{equation}
which can be solved analytically:
\begin{equation}
\label{pertsol1}
\phi_{\tilde k}(\eta) = C_1 F(\eta) + C_2 F(\eta)^\ast\,\,,
\end{equation}
where
\begin{equation}
\label{pertsol2}
F(\eta) = \left(\frac{\sqrt{A}}{2} + i A \sqrt{\eta}\right)\exp(-2i
\sqrt{A \eta}) \,\,.
\end{equation}
The two constants can be specified in formal analogy with the standard
procedure by picking the positive ``frequency'' branch and normalizing
according to the Wronskian condition. The result is regular at $\eta =
0$.  A preliminary analysis appears to indicate that there exists a unique solution
for which  $\phi^\dagger \phi$ is analytic at creation time and that
it corresponds to this solution. If this solution indeed
evolves into the later adiabatic vacuum solution then this would be a
desirable intrinsic mechanism for fixing the
vacuum \footnote{The authors of Ref.~\cite{Eea01}, which appeared after this work was
first posted, use Eqs.~(\ref{pertsol1},\ref{pertsol2}) as the leading term in the
initial conditions of a numerical analysis. Setting $C_2=0$, they
reproduce the same perturbation amplitude as the one predicted by
starting in the adiabatic vacuum, providing some support for our conjecture.}.

\begin{acknowledgments}
JCN is would like to thank Renaud Parentani for illuminating discussions.
\end{acknowledgments}

%\bibliography{../../input/bib/early_universe}

\begin{thebibliography}{10}
\expandafter\ifx\csname bibnamefont\endcsname\relax
  \def\bibnamefont#1{#1}\fi
\expandafter\ifx\csname bibfnamefont\endcsname\relax
  \def\bibfnamefont#1{#1}\fi
\expandafter\ifx\csname url\endcsname\relax
  \def\url#1{\texttt{#1}}\fi
\expandafter\ifx\csname urlprefix\endcsname\relax\def\urlprefix{URL }\fi
\expandafter\ifx\csname bibinfo\endcsname\relax \def\bibinfo#1#2{#2}\fi
\expandafter\ifx\csname eprint\endcsname\relax \def\eprint#1{#1}\fi

\bibitem{LL00}
\bibinfo{author}{\bibfnamefont{A.~R.} \bibnamefont{Liddle}} \bibnamefont{and}
  \bibinfo{author}{\bibfnamefont{D.~H.} \bibnamefont{Lyth}},
  \emph{\bibinfo{title}{Cosmological inflation and large-scale structure}}
  (\bibinfo{publisher}{Univ. Pr.}, \bibinfo{address}{Cambridge, UK},
  \bibinfo{year}{2000}).

\bibitem{U81}
\bibinfo{author}{\bibfnamefont{W.~G.} \bibnamefont{Unruh}},
  \bibinfo{journal}{Phys. Rev. Lett.} \textbf{\bibinfo{volume}{46}},
  \bibinfo{pages}{1351} (\bibinfo{year}{1981}).

\bibitem{J91}
\bibinfo{author}{\bibfnamefont{T.}~\bibnamefont{Jacobson}},
  \bibinfo{journal}{Phys. Rev.} \textbf{\bibinfo{volume}{D44}},
  \bibinfo{pages}{1731} (\bibinfo{year}{1991}).

\bibitem{U95}
\bibinfo{author}{\bibfnamefont{W.~G.} \bibnamefont{Unruh}},
  \bibinfo{journal}{Phys. Rev.} \textbf{\bibinfo{volume}{D51}},
  \bibinfo{pages}{2827} (\bibinfo{year}{1995}).

\bibitem{BMPe95}
\bibinfo{author}{\bibfnamefont{R.}~\bibnamefont{Brout}},
  \bibinfo{author}{\bibfnamefont{S.}~\bibnamefont{Massar}},
  \bibinfo{author}{\bibfnamefont{R.}~\bibnamefont{Parentani}},
  \bibnamefont{and} \bibinfo{author}{\bibfnamefont{P.}~\bibnamefont{Spindel}},
  \bibinfo{journal}{Phys. Rev.} \textbf{\bibinfo{volume}{D52}},
  \bibinfo{pages}{4559} (\bibinfo{year}{1995}), \eprint{hep-th/9506121}.

\bibitem{CJ96}
\bibinfo{author}{\bibfnamefont{S.}~\bibnamefont{Corley}} \bibnamefont{and}
  \bibinfo{author}{\bibfnamefont{T.}~\bibnamefont{Jacobson}},
  \bibinfo{journal}{Phys. Rev.} \textbf{\bibinfo{volume}{D54}},
  \bibinfo{pages}{1568} (\bibinfo{year}{1996}), \eprint{hep-th/9601073}.

\bibitem{BGLe99}
\bibinfo{author}{\bibfnamefont{R.}~\bibnamefont{Brout}},
  \bibinfo{author}{\bibfnamefont{C.}~\bibnamefont{Gabriel}},
  \bibinfo{author}{\bibfnamefont{M.}~\bibnamefont{Lubo}}, \bibnamefont{and}
  \bibinfo{author}{\bibfnamefont{P.}~\bibnamefont{Spindel}},
  \bibinfo{journal}{Phys. Rev.} \textbf{\bibinfo{volume}{D59}},
  \bibinfo{pages}{044005} (\bibinfo{year}{1999}), \eprint{hep-th/9807063}.

\bibitem{SS00}
\bibinfo{author}{\bibfnamefont{H.}~\bibnamefont{Saida}} \bibnamefont{and}
  \bibinfo{author}{\bibfnamefont{M.}~\bibnamefont{Sakagami}},
  \bibinfo{journal}{Phys. Rev.} \textbf{\bibinfo{volume}{D61}},
  \bibinfo{pages}{084023} (\bibinfo{year}{2000}), \eprint{gr-qc/9905034}.

\bibitem{MB00}
\bibinfo{author}{\bibfnamefont{J.}~\bibnamefont{Martin}} \bibnamefont{and}
  \bibinfo{author}{\bibfnamefont{R.~H.} \bibnamefont{Brandenberger}},
  \bibinfo{journal}{Phys. Rev.} \textbf{\bibinfo{volume}{D63}},
  \bibinfo{pages}{123501} (\bibinfo{year}{2001}), \eprint{hep-th/0005209}.

\bibitem{N00}
\bibinfo{author}{\bibfnamefont{J.~C.} \bibnamefont{Niemeyer}},
  \bibinfo{journal}{Phys. Rev.} \textbf{\bibinfo{volume}{D63}},
  \bibinfo{pages}{123502} (\bibinfo{year}{2001}), \eprint{astro-ph/0005533}.

\bibitem{KG00}
\bibinfo{author}{\bibfnamefont{J.}~\bibnamefont{Kowalski-Glikman}},
  \bibinfo{journal}{Phys. Lett.} \textbf{\bibinfo{volume}{B499}},
  \bibinfo{pages}{1} (\bibinfo{year}{2001}), \eprint{astro-ph/0006250}.

\bibitem{NP01}
\bibinfo{author}{\bibfnamefont{J.~C.} \bibnamefont{Niemeyer}} \bibnamefont{and}
  \bibinfo{author}{\bibfnamefont{R.}~\bibnamefont{Parentani}}
  (\bibinfo{year}{2001}), \eprint{astro-ph/0101451}.

\bibitem{MBP01}
\bibinfo{author}{\bibfnamefont{L.}~\bibnamefont{Mersini}},
  \bibinfo{author}{\bibfnamefont{M.}~\bibnamefont{Bastero-Gil}},
  \bibnamefont{and} \bibinfo{author}{\bibfnamefont{P.}~\bibnamefont{Kanti}}
  (\bibinfo{year}{2001}), \eprint{hep-ph/0101210}.

\bibitem{K00}
\bibinfo{author}{\bibfnamefont{A.}~\bibnamefont{Kempf}},
  \bibinfo{journal}{Phys. Rev. D} \textbf{\bibinfo{volume}{63}},
  \bibinfo{pages}{083514} (\bibinfo{year}{2001}), \eprint{astro-ph/0009209}.

\bibitem{K98}
\bibinfo{author}{\bibfnamefont{A.}~\bibnamefont{Kempf}}
  (\bibinfo{year}{1998}), \eprint{hep-th/9810215}.

\bibitem{W96}
\bibinfo{author}{\bibfnamefont{E.}~\bibnamefont{Witten}},
  \bibinfo{journal}{Phys. Today} \textbf{\bibinfo{volume}{49 (4)}},
  \bibinfo{pages}{24} (\bibinfo{year}{1996}).

\bibitem{K94}
\bibinfo{author}{\bibfnamefont{A.}~\bibnamefont{Kempf}}, \bibinfo{journal}{J.
  Math. Phys.} \textbf{\bibinfo{volume}{35}}, \bibinfo{pages}{4483}
  (\bibinfo{year}{1994}), \eprint{hep-th/9311147}.

\bibitem{K97}
\bibinfo{author}{\bibfnamefont{A.}~\bibnamefont{Kempf}}, \bibinfo{journal}{J.
  Phys.} \textbf{\bibinfo{volume}{A30}}, \bibinfo{pages}{2093}
  (\bibinfo{year}{1997}), \eprint{hep-th/9604045}.

\bibitem{BD84}
\bibinfo{author}{\bibfnamefont{N.~D.} \bibnamefont{Birrell}} \bibnamefont{and}
  \bibinfo{author}{\bibfnamefont{P.~C.~W.} \bibnamefont{Davies}},
  \emph{\bibinfo{title}{Quantum Fields in Curved Space}}
  (\bibinfo{publisher}{Univ. Pr.}, \bibinfo{address}{Cambridge, UK},
  \bibinfo{year}{1984}).

\bibitem{T00}
\bibinfo{author}{\bibfnamefont{T.}~\bibnamefont{Tanaka}}
  (\bibinfo{year}{2000}), \eprint{astro-ph/0012431}.

\bibitem{J00}
\bibinfo{author}{\bibfnamefont{T.}~\bibnamefont{Jacobson}},
  \bibinfo{journal}{Prog. Theor. Phys. Suppl.} \textbf{\bibinfo{volume}{136}},
  \bibinfo{pages}{1} (\bibinfo{year}{2000}), \eprint{hep-th/0001085}.

\bibitem{Eea01}
\bibinfo{author}{\bibfnamefont{R.}~\bibnamefont{Easther}},
  \bibinfo{author}{\bibfnamefont{B.~R.} \bibnamefont{Greene}},
  \bibinfo{author}{\bibfnamefont{W.~H.} \bibnamefont{Kinney}},
  \bibnamefont{and} \bibinfo{author}{\bibfnamefont{G.}~\bibnamefont{Shiu}}
  (\bibinfo{year}{2001}), \eprint{hep-th/0104102}.

\end{thebibliography}

\end{document}